\documentclass[showpacs,twocolumn,aps]{revtex4}
\usepackage{epsfig}

\begin{document}
\title{Correlations in Ballistic Processes} 
\author{E.~Trizac}
\email{Emmanuel.Trizac@th.u-psud.fr} 
\affiliation{Laboratoire de Physique Th\'eorique 
(UMR 8627 du Cnrs), B\^atiment 210, 
Universit\'e Paris-Sud, 91405 Orsay, France}
\author{P.~L.~Krapivsky}
\email{paulk@bu.edu} \affiliation{Center for Polymer Studies, and
Department of Physics, Boston University, Boston, MA 02215, USA}
\begin{abstract}
  We investigate a class of reaction processes in which particles move
  ballistically and react upon colliding.  We show that correlations between
  velocities of colliding particles play a crucial role in the long time
  behavior. In the reaction-controlled limit when particles undergo mostly
  elastic collisions and therefore are always near equilibrium, the
  correlations are accounted analytically. For ballistic aggregation, for
  instance, the density decays as $n\sim t^{-\xi}$ with $\xi=2d/(d+3)$ in the
  reaction-controlled limit in $d$ dimensions, in contrast with well-known
  mean-field prediction $\xi=2d/(d+2)$.
\end{abstract}
\maketitle

Ballistic-controlled reaction processes
\cite{CPY,Jiang,brl,mp,bklr,bkr,emm,F,PRE,Ley} exhibit rich atypical behaviors,
e.g. the persistent dependence of decay exponents on the spatial dimension
$d$ implying absence of the upper critical dimension; not surprisingly,
ballistic-controlled processes proved very challenging to theoretical
treatments.  The key such process is ballistic aggregation \cite{CPY} in
which particles merge upon collisions so that mass and momentum are conserved
(energy is necessarily lost).  This model arises in various contents, e.g.,
it mimics the merging of coherent structures (like vortices or thermal
plumes) and accumulation of cosmic dust into planetesimals \cite{Z}. The
one-dimensional (1D) version has also an interesting connection with dynamics
of shocks representing solutions of the inviscid Burgers equation \cite{B,K}.
Ballistic aggregation was first investigated in a pioneering paper \cite{CPY}
by Carnevale, Pomeau, and Young who argued that basic physical quantities
behave algebraically in the long time limit, e.g., the density decays as
\begin{equation}
\label{nt}
n(t)\sim t^{-\xi}, \qquad \xi=\frac{2d}{d+2}
\end{equation}
in $d$ dimensions. To understand this result, one can use \cite{bklr} a rate
equation $dn/dt=-n/\tau$ for the density.  The mean time $\tau$ between
collisions related to the root mean squared (rms) velocity $V$, radius $R$,
and density through $n V\tau R^{d-1}\sim 1$. Mass conservation implies that
the average mass is $m\sim n^{-1}$. Therefore $R\sim n^{-1/d}$ and
\begin{equation}
\label{nt-eq}
\frac{dn}{dt}=-n^2 V R^{d-1}=-n^{1+1/d}\,V.
\end{equation}
The particle of mass $m$ is formed from $m$ original particles (we measure
mass in units of the initial mass and velocity in units of the initial rms
velocity). Assuming velocities of those original particles uncorrelated we
find that the average momentum $p$ and velocity $V$ scale as
\begin{equation}
\label{wrong}
p\sim m^{1/2}, \qquad V=p/m\sim n^{1/2}.
\end{equation}
Plugging (\ref{wrong}) into (\ref{nt-eq}) and solving for $n(t)$ yields
(\ref{nt}).

Surprisingly the prediction $\xi=2d/(d+2)$ for the decay exponent --- perhaps
the most known result in the field of ballistic-controlled processes --- is
{\em erroneous}.  It turns out that the mean-field assumption that velocities
of original particles contained within a typical aggregate particle are
uncorrelated is incorrect in any finite dimension --- only when $d\to\infty$
and velocities are orthogonal to each other with probability one, they are
indeed uncorrelated.  The failure of the mean-field no-correlation assumption
(\ref{wrong}) has not been appreciated because the resulting formula
$\xi_d=2d/(d+2)$ is correct both for $d=1$ and $d=\infty$.  (No trivial
explanation of the former assertion is known yet the relation to the Burgers
equation via the particles$\Longleftrightarrow$shocks mapping \cite{B,K} and
the $t^{2/3}$ growth of the separations between adjacent shocks established
by Burgers many years ago \cite{B} prove that $\xi_1=2/3$).  Since $\xi_d$
monotonously increases with $d$, it is not surprising that the actual values
are not so different from the mean-field prediction (\ref{nt}).  Therefore
the observed disagreement in two dimensions \cite{emm} could be attributed to
insufficient scale of the simulations. Interestingly, the beauty of ballistic
aggregation in 1D where the model admits an exact solution \cite{mp,F} and
exhibits a deep connection to the Burgers equation has supported the
incorrect prediction (\ref{nt}) in higher dimensions.

The purpose of this article is twofold. First, we clarify the role of
velocity correlations in the general case, where they lead to significant
deviations from mean-field predictions.  Second, we propose a procedure that
allows an analytical treatment of correlations for virtually any ballistic
reaction process in the reaction-controlled limit; in particular, this
method gives exact decay exponents.

%% We start by showing the importance of velocity correlations on a simple toy
%% model, where the connection between the standard mean-field procedure and a
%% simplified kinetic theory approach (namely the Maxwell model) becomes
%% transparent.  In most cases, the analytical computation of decay exponents is
%% intractable and the more subtle case of ballistic aggregation is revisited by
%% means of a powerful technique providing the first accurate exponents for this
%% model.  We finally consider a reaction controlled limit where those
%% correlations may be accounted for analytically for virtually any ballistic
%% reaction process.  This leads to exact predictions for the scaling exponents.

The no-correlation assumption is generally wrong for all ballistic-controlled
processes, so we first demonstrate this assertion for one particularly simple
process.  We choose a toy ballistic aggregation model in which all particles
are identical and when two particles moving with velocities ${\bf v}_1$ and
${\bf v}_2$ collide, they form an aggregate particle moving with velocity
${\bf v}={\bf v}_1+{\bf v}_2$.  Compared to the original ballistic
aggregation model, the toy model has a number of advantageous properties.
First, the volume fraction decays indefinitely thereby driving the system
into the dilute limit and justifying ignoring multiple collisions. Second,
the mean free path $n^{-1}$ grows {\em faster} than the inter-particle
distance $n^{-1/d}$ for $d>1$.  These two features indicate that for $d>1$
the Boltzmann equation approach is exact at large times.

For the toy model, (\ref{nt-eq}) becomes $dn/dt=-n^2 V$ and the supposed
absence of correlations gives $V\sim n^{-1/2}$. Thus the mean-field argument
implies $n\sim t^{-\xi}$ with $\xi=2$ independently on dimension $d$.
Numerically we find that this universality does {\em not} hold: $\xi$
increases with dimension and approaches the mean-field prediction only when
$d\to\infty$. For instance, we find (with an accuracy better than 1\%)
\begin{equation}
\label{xi-toy}
\xi \approx
\cases{
1.33& when\quad $d=1$,\cr
1.55& when\quad $d=2$,\cr
1.65& when\quad $d=3$.}
\end{equation}
These results were obtained by solving 
the Boltzmann equation describing the toy model 
\begin{eqnarray}
\label{BE}
\frac{\partial P({\bf v},t)}{\partial t}
\!\!&=&\!\!\!\!\int d{\bf u}\,d{\bf w}\,P({\bf u},t)\,P({\bf w},t)
|{\bf u}-{\bf w}|\,
\delta({\bf u}+{\bf w}-{\bf v})\nonumber\\
&-&2P({\bf v},t)\int d{\bf w}\,P({\bf w},t) |{\bf v}-{\bf w}|\,.
\end{eqnarray}
We have solved this equation numerically implementing a Direct Monte Carlo
(DMC) simulation scheme (see e.g. \cite{Bird} for the general method, and
\cite{PRE} for an application to a ballistic-controlled reaction process).
The idea is to rephrase Eq.~(\ref{BE}) as a stochastic process. In each step
two particles, say with velocities ${\bf u}$ and ${\bf w}$, are selected at
random among a population of $N$ particles, and the reaction happens with a
probability proportional to $|{\bf u}-{\bf w}|$. If the reaction has been
accepted, a new particle of velocity ${\bf u}+{\bf w}$ replaces two original
particles, so the number of particles changes to $N-1$.  The time is
incremented by $(N^2 |{\bf u}-{\bf w}|)^{-1}$, and the process is iterated
again. This numerical scheme allows us to treat systems with initial number
of particles of the order of $10^7$.  The master equation associated to this
Markov chain is precisely (\ref{BE}) so that we obtain the numerically exact
solution of our problem.  The exponent values (\ref{xi-toy}) significantly
differ from the mean-field prediction $\xi=2$, and leave no doubt that the
no-correlation assumption is wrong.

In the Boltzmann equation (\ref{BE}), the relative velocity $|{\bf v}-{\bf
  w}|$ gives the rate of collisions and its non-linear character makes
analytical progress hardly possible.  An old trick to overcome this
difficulty is to replace the actual relative velocity by the rms velocity
\cite{max}. This results in the Maxwell model that played an important role
in the development of kinetic theory \cite{classical,e}.  For the toy model,
we have (hereafter the dependence on time is suppressed for ease of notation)
\begin{equation}
\label{BM}
\frac{1}{V}\,\frac{\partial P({\bf v})}{\partial t}
=\int d{\bf u}\,P({\bf u})\,P({\bf v}-{\bf u})-2nP({\bf v})\,.
\end{equation}
Integrating (\ref{BM}) we find that the density $n=\int d{\bf w}\, P({\bf
  w})$ satisfies $dn/dt=-n^2 V$ while $n V^2 = \int d{\bf w}\, w^2 P({\bf
  w})$ remains constant. Hence $V=n^{-1/2}$ and $\xi=2$ showing that the
mean-field no-correlation approach is essentially the Maxwell model in
context of ballistic processes \cite{bk}.  The Maxwell model is an
uncontrolled approximation 
to the Boltzmann equation for the hard sphere gas
and, not surprisingly, the exponents found within this approach are generally
erroneous (see \cite{Rqvhp} for an alternative simplification,
the so-called very hard particle approach).
Of course, one could anticipate that the exponent $\xi=2$
characterizes the Maxwell model without computations --- the essence of the
Maxwell model, that is the fact that collisions are completely random,
assures that the no-correlation condition does hold.  
%The Maxwell model is in fact so simple that in addition to the exponent $\xi$, 
%the exact velocity distribution can be determined analytically \cite{prep}.

We now present an argument that emphasizes the role and importance of
correlations between velocities of colliding particles and applies to all
ballistic-controlled reaction processes.  The key point is to supplement
an evolution equation for the mass density by an evolution equation for the
density of kinetic energy. For an arbitrary ballistic-controlled reaction
process we denote $P(m,{\bf v},t)$ the joint mass-velocity distribution
function, and $e=mv^2$ the kinetic energy of a given particle (for the toy
model, we set $m=1$).  The evolution equations for the density $n$ and
kinetic energy density $n E = \int m v^2 P(m,{\bf v},t)\,dm\, d{\bf v} = n
\langle m v^2\rangle$ read
\begin{eqnarray}
\label{ne}
\frac{d n}{dt} = - \frac{n}{\tau}\,,\qquad
\frac{d (n E)}{dt} = - 
\frac{n \langle \Delta e\rangle_{\hbox{\scriptsize coll}}}{\tau}.
\label{eq:alpha}
\end{eqnarray}
The first equation is just the definition of the time dependent collision
frequency $1/\tau$, the second additionally contains the kinetic energy
$\langle \Delta e\rangle_{\hbox{\scriptsize coll}}$ lost on average in a
binary collision. In the scaling regime the quantities $\langle \Delta
e\rangle_{\hbox{\scriptsize coll}}$ and $E = \langle m v^2 \rangle$ exhibit
the same time dependence, so the dissipation parameter $\alpha = \langle
\Delta e\rangle_{\hbox{\scriptsize coll}}/E$ is asymptotically time
independent.  {}From Eqs.~(\ref{ne}) we get $d\ln(nE)/d\ln n = \alpha$, or
$V^2=nE\sim n^\alpha$. The mean free path argument $\tau^{-1} \sim n V
R^{d-1} \sim t^{-1}$ gives $n^{1/d}V\sim t^{-1}$ for ballistic aggregation.
Combining these two relations and the definition of $\xi$ we obtain $\xi =
(1/d+\alpha/2)^{-1}$. Similarly for the toy model $nV^2\sim n^\alpha$ and
$nV\sim t^{-1}$ leading to $\xi = 2/(1+\alpha)$.

To use this formalism, we must precisely define the collisional average
involved in (\ref{eq:alpha}). An average change of a quantity ${\cal A}(1,2)$
in a binary collision is \cite{radius}
\begin{equation}
\langle \Delta {\cal A}\rangle_{\hbox{\scriptsize coll}} = 
\frac{\int\! d1 \,d 2\, |{\bf v}_1-{\bf v}_2|^\nu\, [\Delta{\cal A}(1,2)]\,
 P(1) \,P(2)}{\int d1 \,d2 \,|{\bf v}_1-{\bf v}_2|^\nu \,
 P(1)\,P(2)},
\label{eq:colldef} 
\end{equation}
where we have used shorthand notations $i=(m_i,{\bf v}_i)$ and $di=d m_i\, d
{\bf v}_i$ ($i=1,2$).  In the key case of hard spheres we have $\nu=1$,
whereas the cases $\nu=0$ and $\nu=2$ correspond the Maxwell and very hard
particle models \cite{Rqvhp}, respectively. We now illustrate the formalism for the toy
model. The kinetic energy lost in a collision is $\Delta e = {\bf v}_1^2+{\bf
  v}_2^2-({\bf v}_1+{\bf v}_2)^2 = -2 \,{\bf v}_1\cdot {\bf v}_2$. Hence 
\begin{eqnarray*}
\langle \Delta e\rangle_{\hbox{\scriptsize coll}} = -2\,
\frac{\int d{\bf v}_1\,d{\bf v}_2\, |{\bf v}_1-{\bf v}_2|^\nu 
({\bf v}_1\cdot {\bf v}_2)\,
 P({\bf v}_1)\,P({\bf v}_2)}
{\int d{\bf v}_1\,d{\bf v}_2\,|{\bf v}_1-{\bf v}_2|^\nu 
P({\bf v}_1)\,P({\bf v}_2)}\,. 
\end{eqnarray*}
For the Maxwell model ($\nu=0$), the isotropy of $P({\bf v},t)$ shows that
$\langle \Delta e\rangle_{\hbox{\scriptsize coll}}=0$, so $\alpha = 0$ and
$\xi=2/(1+\alpha)=2$ in agreement with our previous calculation.  Similarly
for very hard particles ($\nu=2$, see \cite{Rqvhp}) 
we use isotropy to simplify $\alpha$ and
arrive at
\begin{eqnarray*}
\alpha= 2\,
\frac{\int d{\bf v}_1\,d{\bf v}_2\, ({\bf v}_1\cdot {\bf v}_2)^2\,
 P({\bf v}_1)\,P({\bf v}_2)}
{\left[\int d{\bf v}\,{\bf v}^2\, P({\bf v})\right]^2}\,. 
\end{eqnarray*}
The isotropy allows to compute the ratio of the integrals to yield
$\alpha=2/d$ leading to $\xi=2d/(d+2)$.  For other values of $\nu$, including
the case of interest $\nu=1$, the dissipation parameter $\alpha$ depends on
details of the velocity distribution, and isotropy alone is not sufficient to
determine $\alpha$.  The reason for the failure of the mean-field argument
--- which amounts to the complete neglect of collisional correlations
($\langle {\bf v}_1\cdot{\bf v}_2\rangle_{\hbox{\scriptsize coll}}=0$) --- is
now clear: in general, a collision involving a pair $({\bf v}_1,{\bf v}_2)$
with a {\em negative} product ${\bf v}_1\cdot{\bf v}_2 < 0$ has a higher
probability than a collision with ${\bf v}_1\cdot{\bf v}_2 > 0$. The
dissipation parameter $\alpha$ is therefore positive so that $\xi =
2/(1+\alpha)<2$.  Thus the mean-field prediction $\xi=2$ is an upper bound
for $\xi$.

 \begin{center}
 \begin{figure}[htb]
% \vspace{4cm}
%\epsfig{figure=./cpy2d.eps,width=7cm,angle=0}
\epsfig{figure=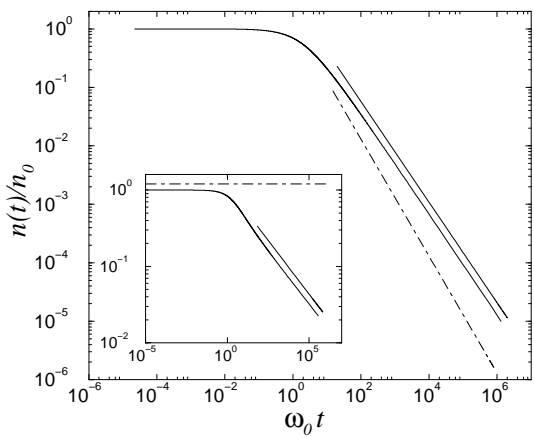,width=7cm,angle=0}
 \caption{Density versus time in the 2D aggregation 
   model.  The continuous straight line has slope $-0.86$.  The inset shows
   that the average energy $\langle m v^2\rangle$ decays as $t^{-0.28}$.
   Mean-field predictions are shown by the dashed lines (slope $-1$ in the
   main graph and 0 in the inset).  The non-linear Boltzmann equation
   describing ballistic aggregation has been solved by DMC for a system of
   $N=4\times 10^7$ particles.  The initial density is $n_0$ and $\omega_0$
   denotes the initial collision frequency of the equilibrium hard sphere
   fluid.
 \label{fig:cpy2d}}
 \end{figure}
 \end{center}

 The above framework applies to any irreversible process with ballistic
 transport. For ballistic aggregation, the omission of collisional
 correlations amounts to setting $\alpha=1$, i.e.  that the typical energy
 dissipated in a collision is the mean kinetic energy per particle. However,
 particles with larger velocities undergo more frequent collisions so that
 the mean energy dissipated exceeds the energy of a typical particle.  Hence
 $\alpha = \langle \Delta e\rangle_{\hbox{\scriptsize coll}}/E >1$ so that
 $\xi$ is smaller than the mean-field prediction $2d/(d+2)$.  Previous
 Molecular Dynamics (MD) simulations have shown that $\xi \simeq 0.85 \pm
 0.04$ in 2D for low volume fractions, with scaling laws extending over 2
 decades in time \cite{emm}. The DMC technique allows to reach much larger
 time scales.  Figure \ref{fig:cpy2d} shows that after an initial transient,
 the density exhibits a clear power law behavior over 5 decades in time.  We
 estimate $\xi \simeq 0.86 \pm 0.005$, in agreement with MD simulations. The
 inset displays the behavior of $E = \langle m v^2\rangle$, the quantity that
 is (asymptotically) time independent according to the mean-field prediction
 (\ref{wrong}); we find $E\sim t^{-0.28}$ \cite{R}.  We have also performed
 DMC and MD simulations in 3D giving $\xi \simeq 1.06\pm 0.01$.  As expected,
 the actual values of $\xi$ are smaller than the mean-field prediction
 $\xi=2d/(d+2)$\cite{note1}.

Ballistic-controlled processes are generally intractable analytically.
Following the fruitful line of attack on difficult problems --- generalize
them! --- let us consider a process in which colliding particles react with
probability $\epsilon$ and scatter elastically with complementary probability
$1-\epsilon$. The mean-field no-correlation argument is so general that it
applies to these processes; in particular, according to mean-field the
exponent $\xi$ is independent on $\epsilon$.  Remarkably, we can now compute
the exponent $\xi$ for one special value of $\epsilon$, viz. for $\epsilon\to
0^+$.  In this {\em reaction-controlled} limit particles undergo mostly elastic
collisions.  Therefore, the particles are always at equilibrium, i.e., the
velocity distribution is Maxwellian. This key feature makes the problem
tractable.  Consider for instance the toy model.  One can compute
\begin{eqnarray*}
\langle \Delta e\rangle_{\hbox{\scriptsize coll}} = -2\,
\frac{\int d{\bf v}_1\,d{\bf v}_2\,|{\bf v}_1-{\bf v}_2|^\nu 
({\bf v}_1\cdot {\bf v}_2)\,
 P({\bf v}_1)\,P({\bf v}_2)}
{\int d{\bf v}_1\,d{\bf v}_2\,|{\bf v}_1-{\bf v}_2|^\nu 
P({\bf v}_1)\,P({\bf v}_2)} 
\end{eqnarray*}
for arbitrary $\nu$ when $P({\bf v})$ is Maxwellian \cite{help}. In
particular for the important case $\nu=1$ we obtain $\alpha=1/d$, so that
$\xi = 2d/(d+1)$. This {\em exact} result provides a useful check of
numerical scheme (our DMC simulations in two and three dimensions are indeed
in excellent agreement with the theoretical prediction).  In contrast, the
mean-field no-correlation argument predicts $\xi = 2$ irrespective of the
value of $\epsilon$. We see again that this is correct only in the $d\to
\infty$ limit.  Note also that $\xi = 2d/(d+1)$ which is exact for the
reaction-controlled ($\epsilon \to 0^+$) version of the toy model provides a
better `guess' for $\xi$ in the original ($\epsilon=1$) model than the
mean-field approach [compare $\xi=1$, 4/3 and 3/2 in 1D, 2D, and 3D to the
numerical values (\ref{xi-toy})].

Remarkably, the exponent $\xi$ in the reaction-controlled limit of ballistic
aggregation can be computed even though the mass distribution $\Pi(m) =
n^{-1}\int d{\bf v}\, P(m,{\bf v})$ is unknown.  The important point is that
when $\epsilon \to 0^+$, the joint mass/kinetic energy distribution function
factorizes.  Then one finds $\alpha=1+1/d$, or equivalently $\xi = 2d/(d+3)$
independently on $\Pi(m)$ \cite{help2}.  This exact result of course agrees
with DMC simulations. Interestingly, it also provides a reasonable
approximation of $\xi$ for the original ($\epsilon=1$) aggregation model:
$\xi=0.8$ in 2D and 1 in 3D, to be compared to 0.86 and 1.06, respectively.

Many other ballistic-reaction processes are solvable in the
reaction-controlled limit. For instance for ballistic annihilation
\cite{brl}, there is no exact solution in any dimension yet in
reaction-controlled limit, the exact value of the density decay exponent is
given by $\xi=4d/(4d+1)$. This result is in surprisingly good agreement with
numerical values for $\epsilon=1$: $\xi=4/5$ vs. 0.804 \cite{Ely} in 1D;
$\xi=8/9 \simeq 0.89$ vs. 0.87 \cite{PRE} in 2D; $\xi = 12/13\simeq 0.92$
against 0.91 \cite{PRE} in 3D. We have studied several other
ballistic-reaction processes \cite{prep}, e.g., a simplified ballistic
aggregation model in which mass and momentum are conserved yet the radius
does not grow. For this model, the mean-field prediction is $\xi=2/3$
independently on dimension $d$, whereas in the reaction-controlled limit, we
get the exact result $\xi= 2d/(3d+1)$ (i.e. 0.571 in 2D and 0.6 in 3D).  It
is again instructive to compare these values with numerical results for
$\epsilon=1$: $\xi\simeq 0.60$ in 2D, and $\xi\simeq 0.62$ in 3D.

We have shown that correlations between velocities of colliding particles 
% --- that have been overlooked so far ---
govern the behavior of all reacting processes with ballistic transport. We
illustrated importance of correlations on several models and demonstrated
that ignoring correlations is equivalent to using the Maxwell model, which is
an uncontrolled approximation of the hard-sphere gas.
%In the reaction-controlled limit when particles undergo mostly elastic collisions 
%and therefore are always near equilibrium, we determined correlations analytically. 
We also devised a procedure that clarifies the role of correlations in the
general case and allows an exact computation of decay exponents in the
reaction-controlled limit, when particles undergo mostly elastic collisions
and therefore are always near equilibrium.  The failure of mean-field theory
to describe this limit emphasizes inevitable presence of correlations in all
reacting processes with ballistic transport.


\begin{thebibliography}{99}
\bibitem{radius} 
     If particles have different radii $R_i$ (in ballistic
      aggregation, $R\sim m^{1/d}$), the collision kernel in
      Eq.~(\ref{eq:colldef}) becomes $(R_1+R_2)^{d-1} |{\bf v}_1-{\bf v}_2|$
      for $\nu=1$, instead of $|{\bf v}_1-{\bf v}_2|$ that is only valid when
      all particles have the same size.
\bibitem{note1}
We also estimated $\xi$ directly from $\alpha$ that was computed by 
measuring the kinetic energy dissipated in successive collisions.
Such a route gives accurate values of $\alpha$ from which we deduced the
error bars mentioned above for $\xi=(1/d+\alpha/2)^{-1}$.  We also
mention that the asymptotic regime does not depend on the initial
conditions chosen.
\end{thebibliography}
\end{document}